\documentclass[pre,twocolumn,floatfix,showpacs,superscriptaddress,
amsmath,amssymb]{revtex4}

\usepackage{graphicx}

\begin{document}

\newcommand{\be}{\begin{equation}}
\newcommand{\bea}{\begin{eqnarray}}
\newcommand{\ee}{\end{equation}}
\newcommand{\eea}{\end{eqnarray}}
\newcommand{\bse}{\begin{subequations}}
\newcommand{\ese}{\end{subequations}}

%\draft{Preprint. Submitted to Physical Review E}

\title{Robustness and Enhancement of Neural Synchronization \\
 by Activity-Dependent Coupling}

\author{V.P.~Zhigulin}
\email{zhigulin@caltech.edu}
\affiliation{Department of Physics,
MC 103-33, California Institute of Technology, Pasadena, CA 91125}
\affiliation{Institute for Nonlinear Science, University of
California, San Diego, La Jolla, CA  92093-0402}

\author{M.I.~Rabinovich}
\affiliation{Institute for Nonlinear Science, University of
California, San Diego, La Jolla, CA  92093-0402}

\author{R.~Huerta} 
\affiliation{Institute for Nonlinear Science, University of
California, San Diego, La Jolla, CA  92093-0402}
\affiliation{GNB, E.T.S. de Ingenier\'{\i}a Inform\'{a}tica,
Universidad Aut\'{o}noma de Madrid, 28049 Madrid (SPAIN)}

\author{H.D.I.~Abarbanel}
\affiliation{Institute for Nonlinear Science, University of
California, San Diego, La Jolla, CA  92093-0402}
\affiliation{Department of Physics and
Marine Physical Laboratory (Scripps Institution of Oceanography),
University of California, San Diego, La Jolla, CA 93093-0402}

%\date{\today}

\begin{abstract}

We study the synchronization of two model neurons
coupled through a synapse having an activity-dependent strength. Our 
synapse follows the rules of Spike-Timing Dependent Plasticity (STDP). 
We show that this plasticity of the coupling between neurons produces 
enlarged frequency locking zones and results in synchronization that 
is more rapid and much more robust against noise than
classical synchronization arising from connections with constant
strength. We also present a simple discrete map model that 
demonstrates the generality of the phenomenon. 

\end{abstract}

\pacs{05.45.Xt, 87.18.Sn, 87.18.Bb}

\maketitle

Synchronous activity among individual neurons or their ensembles
is a robust phenomenon observed in many regions of the brain, in
sensory systems and in other neural networks. With constant
synaptic connections the regions of neural synchronization are quite
narrow in parameter space and the origin of the observed robustness 
of synchronization is not clear. It is known that many neurons in 
the cortex, in the cerebellum and in other neural systems are coupled
through excitatory synaptic connections whose strength can be altered
through activity-dependent plasticity. Indeed, this plasticity is 
widely thought to underlie learning processes, and in itself 
constitutes a broadly interesting phenomenon. Here we discuss 
its role in the synchronization of neurons in a network.

There have been recent experimental advances in the understanding of 
such plasticity, and, in particular, of the critical dependence on 
timing in presynaptic and postsynaptic signaling. Two manifestations
of this kind of synaptic plasticity are the Spike-Timing Dependent 
Plasticity (STDP)~\cite{Markram97,Bi98} seen in excitatory connections
between neurons, and its inverse, observed, for example, in the 
connections between excitatory and inhibitory neurons in the 
electrosensory lobe of fish~\cite{Bell99}. The connections between 
excitatory neurons through inhibitory interneurons are typical in 
sensory systems~\cite{Laurent99,Rabin00} and cerebral 
cortex~\cite{Bear00}. These also express synaptic plasticity~\cite{Perez01}
and play an important role in the control and synchronization of neural
ensembles in hippocampus.

We report here on the synchronization of two model 
neurons coupled through STDP or inverse STDP
synapses. We demonstrate that such coupling leads to neural
synchronization which is more rapid, more flexible and much more 
robust against noise than synchronization mediated by constant 
strength connections. 
(For reviews, see~\cite{Glass01,Elson98,Coombes99}). 
We also build a simple discrete map that illustrates the 
enhancement of synchronization by activity-dependent coupling.
The map allows us to speculate about the general applicability 
of learning-enhanced synchronization.

We consider here the simplest neural network: two neurons with
unidirectional, activity-dependent excitatory synaptic coupling. 
Each neuron is described by the Hodgkin-Huxley equations
with standard Na, K, and `leak' currents~\cite{Traub91}:
\bea
C\frac{dV_{i}(t)}{dt} &=& -g_{Na} m_i(t)^3 h_i(t)(V_i(t)-E_{Na})\nonumber \\
&-& g_K n_i(t)^4(V_i(t)-E_K)-g_L(V_{i}(t)-E_L) \nonumber \\
&-&I_{syn}(t)+I_{stim},
\label{eq1} 
\eea
where $i=1,2$. 

Each of the activation and inactivation variables 
$y_i(t) = \{n_i(t),m_i(t),h_i(t)\}$ satisfies first-order kinetics
\be
\frac{dy_i(t)}{dt}=\alpha_y (V_i(t))(1-y_i(t))-\beta_y(V_i(t)) y_i(t).
\ee 
The parameters in these equations are given in~\cite{params}.

Each neuron receives a constant input $I_{stim}$ forcing it to spike 
with a constant, $I_{stim}$--dependent frequency. The second neuron is
synaptically driven by the first via an excitatory current dependent on
the postsynaptic $V_2(t)$ and presynaptic $V_1(t)$ membrane voltages:
\be
I_{syn}(t) = g(t)S(t)V_{2}(t).
\ee
$S(t)$ is the fraction of open synaptic channels. It satisfies
first-order kinetics:
\be
\frac{dS(t)}{dt} = \alpha (1 - S(t))H(V_{1}(t)) - \beta S(t),
\ee
with  $H(V_{1}(t)) = (1 + \tanh(10 V_{1}(t)))/4$. 

The time dependent synaptic coupling strength $g(t)$ is conditioned 
by the dynamics of the pre- and postsynaptic neurons. We consider 
two types of activity-dependent couplings: (1) an excitatory synapse 
with STDP, and (2) an excitatory synapse with inverse STDP. Through 
STDP $g(t)$ changes by $\Delta g(t)$ which is a function of the time
difference $\Delta t= t_{post}-t_{pre}$ between the times of post- and
presynaptic spikes. We use the additive update rule
\be
\Delta g(t)=G(\Delta t)= A\text{ sgn}(\Delta t) \exp{(-\gamma|\Delta t|)}
\ee
for STDP, and $\Delta g(t) = -G(\Delta t)$ for inverse STDP. We used 
$A = 0.004 \,\mu\,$S and $\gamma = 0.15 \text{ms}^{-1}$. 

\begin{figure}
  \begin{center}
\leavevmode
    \includegraphics[width=8.5 cm]{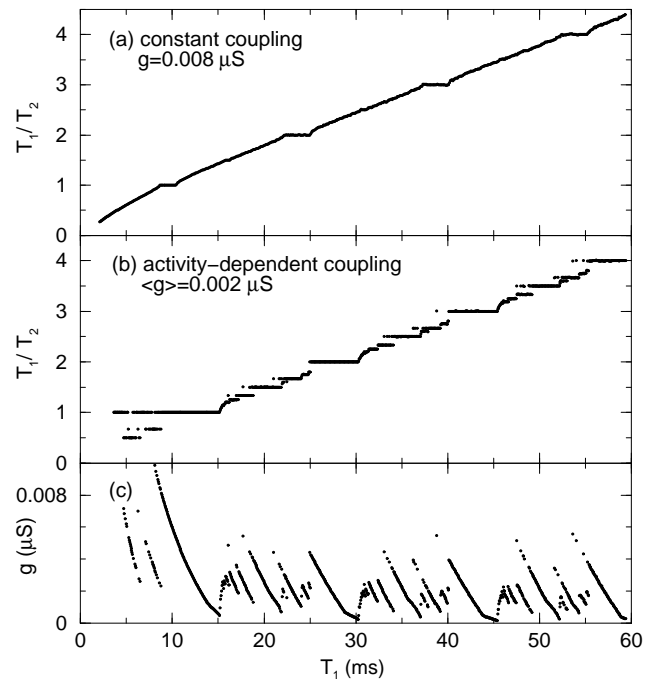}
  \end{center}
\caption{\label{DS_All}Devil's Staircase for (a) constant synaptic 
strength and (b) synaptic strength varying according to inverse STDP 
coupling. $T_1$ and $T_2$ are the observed periods of the presynaptic 
(driving) neuron and postsynaptic (driven) neuron respectively. In (c)
the final value of synaptic strength is displayed.}
\end{figure}

We studied the synchronization properties of this coupled system by 
setting the autonomous period of the postsynaptic neuron to 15 ms, 
then evaluating the actual period of its oscillation $T_2$ as a 
function of the imposed autonomous oscillation period $T_1$ of the 
presynaptic neuron. In Fig.~\ref{DS_All} we show $T_1/T_2$ as a 
function of $T_1$ in two cases: (a) a synaptic coupling with constant
strength $0.008\,\mu S$ and (b) a synaptic coupling with inverse STDP.
In the later case the steady-state coupling strength depends on the 
ratio of neuronal frequencies (c). Its average over all $T_1$ values is 
$0.002\, \mu S$, which is much lower than the strength in the case 
of constant coupling.

In Fig.~\ref{DS_All}a we see the familiar `Devil's Staircase' 
associated with frequency locking domains of a driven nonlinear 
oscillator. Only frequency locking with ratios 1:1, 2:1, 3:1, and 4:1
leads to synchronization plateaus with significant width. 
In Fig.~\ref{DS_All}b we see that the synchronization domains are 
substantially broadened due to activity-dependent coupling, especially
for $T_1/T_2=1$. Some synchronization plateaus exhibit multistability,
which we confirmed by observing the associated hysteresis. These 
results show that even a weak, but adaptive connection
with strength that is determined dynamically is able to greatly 
enhance and enrich synchronization.
%%%%%%%%%%%%%%%%%%%%%%%%%%%%%%%%%%%%%%%%%%%%%%%%%
%%%%%%%%%%%%%%%%%%%%%%%%%%%%%%%%%%%%%%%%%%%%%%%%%%

We also studied the robustness of this enhanced synchronization in the
presence of noise by adding zero mean, Gaussian, white noise to the 
membrane currents of each neuron. We examined the behaviour of the 
system with RMS noise amplitudes $\sigma=0.01,\,0.05,\,0.1,
\text{and }0.5$ nA.

\begin{figure}
  \begin{center}
\leavevmode
    \includegraphics[width=8.5 cm]{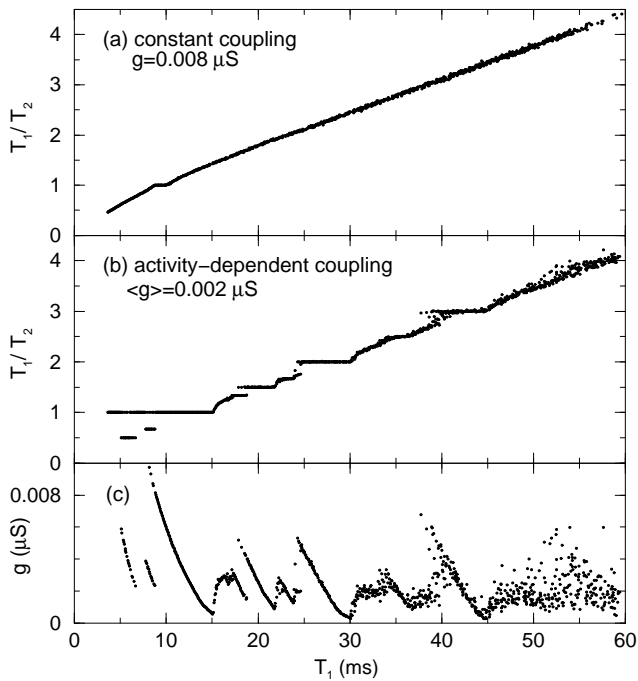}
  \end{center}
\caption{\label{DS_All_ns}Same as Fig.~\ref{DS_All}, but with zero 
mean, Gaussian, white noise with $\sigma=0.1$ nA added to the 
membrane currents.}
\end{figure}

For $\sigma=0.01$ nA no phase-locking plateaus were destroyed. At 
$\sigma=0.05$ nA the 4:1 plateau became distorted. Larger 
$\sigma$ sequentially eliminated synchronization plateaus until only 
the 1:1 plateau remained. The 1:1 plateau was seen for all $\sigma$.
In Fig.~\ref{DS_All_ns} we illustrate the effect of the noise on 
synchronization when $\sigma=0.1$ nA with (a) constant and (b) inverse
STDP coupling. While in (a) most of the plateaus have disappeared, in
(b) the 1:1, the 2:1 and even the 3:1 frequency locking regimes 
remained. In sharp distinction to classical synchronization, frequency
locking through activity-dependent coupling is significantly
more robust in the presence of noise. 
%%%%%%%%%%%%%%%%%%%%%%%%%%%%%%%%%%%%%%%%%%%%%%%%%
%%%%%%%%%%%%%%%%%%%%%%%%%%%%%%%%%%%%%%%%%%%%%%%%%%

\begin{figure}
  \begin{center}
\leavevmode
    \includegraphics[width=8.5 cm]{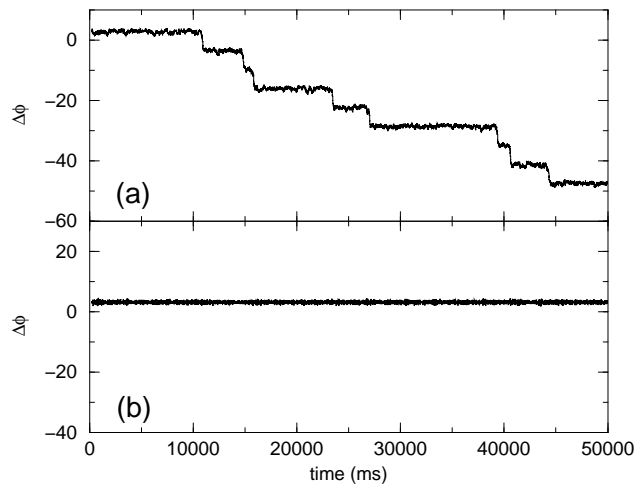}
  \end{center}
\caption{\label{Phases}The difference of oscillation phases of two
neurons as a function of time in the cases of (a) constant and
(b) activity-dependent coupling. }
\end{figure}

To understand the mechanisms behind such a remarkable robustness we
studied the diffusion of oscillation phase caused by noise. For 
$\sigma=0.5$ nA in Fig.~\ref{Phases}a we show that  in the case of 1:1 
synchronization and coupling with constant strength $0.008\, \mu S$ 
noise-induced phase diffusion results in 2$\pi$ phase slips
that destroy synchronized state. Quite contrary Fig.~\ref{Phases}b
shows that in the case of activity-dependent coupling phase slips are absent 
and the phase difference does not increase. In this particular case 
the strength of coupling varied around the mean of $0.0064\, \mu S$ 
with standard deviation of $0.0026\, \mu S$.
%%%%%%%%%%%%%%%%%%%%%%%%%%%%%%%%%%%%%%%%%%%%%%%%%
%%%%%%%%%%%%%%%%%%%%%%%%%%%%%%%%%%%%%%%%%%%%%%%%%%

\begin{figure}
  \begin{center}
\leavevmode
    \includegraphics[width=8.5 cm]{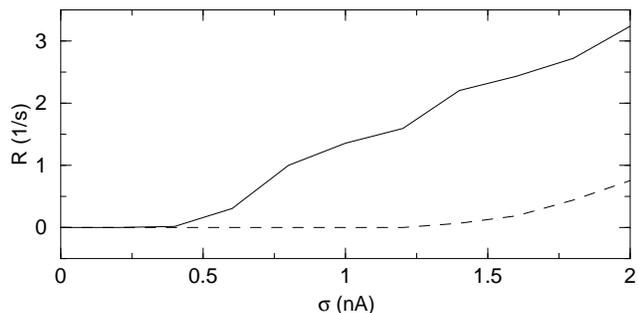}
  \end{center}
\caption{\label{PhaseSlips}Average rate of phase slips as a 
function of RMS noise amplitude for the case of 1:1 synchronization
and constant (solid line) or activity-dependent (dashed line) coupling.}
\end{figure}

In Fig.~\ref{PhaseSlips} we plot the average rate of phase slips 
for different amplitudes of the noise. In line with the above 
observation we see that in the case of activity-dependent coupling
(dashed line) phase slips are suppressed in a wide range of noise 
amplitudes. We argue here that this suppression of phase slips is 
the primary mechanism responsible for robustness of synchronization
mediated by activity-dependent coupling. After the introduction of
a discrete map model we will discuss this mechanism in more detail.     
%%%%%%%%%%%%%%%%%%%%%%%%%%%%%%%%%%%%%%%%%%%%%%%%%
%%%%%%%%%%%%%%%%%%%%%%%%%%%%%%%%%%%%%%%%%%%%%%%%%%

We also considered synchronization 
through an activity-dependent synapse in the interesting case when the
presynaptic neuron produces bursts of spikes and the postsynaptic 
neuron spikes irregularly. We found that synchronization through
an STDP synapse is very fast; even a few spikes are enough for the 
frequency locking to establish itself. Neurons in the same set up
with constant coupling synchronize much more slowly and {\em only} 
if the strength of the connection is appropriate for the given ratio 
of their frequencies. Hence, activity dependent synapses allow 
adaptation `on the run,' synching a postsynaptic neuron to the firing
properties of its presynaptic partner.

To understand the above results in a general way we have constructed a
discrete time map model of periodic generators with STDP-like
coupling. This map accounts for the dependence of the coupling strength
on the activity of generators. Take $T_1^0$ and $T_2^0$ as the autonomous 
periods of the first and second generators. As a result of unidirectional
coupling, the period of the second generator will change by some amount 
$\Delta T$ each time it receives a spike from the first generator.
Assuming initial phases to be 0, the time of the $n+1$-st
spike of the first generator and $m+1$-st spike of the second generator
are taken to satisfy
\bse \label{t-nm} \bea
 t_{n+1}^{(1)} & = & t_{n}^{(1)} + T_1^0
\\t_{m+1}^{(2)} & = & t_{m}^{(2)} + T_2^0 - \Delta T_{m,n},
\eea \ese
where $n$ and $m$ are such that $t_{m}^{(2)} \leq t_{n}^{(1)} \leq
t_{m+1}^{(2)}$. In general, $\Delta T_{m,n}$ would be a
function of $T_1^0$, $T_2^0$, $t_{n}^{(1)}$, $t_{m}^{(2)}$, and
the coupling strength $g_{m,n}$. We argue that the two main
variables here are $t_{n}^{(1)} - t_{m}^{(2)}$, and $g_{m,n}$.
In the simplest case $\Delta T_{m,n}$ can be approximated by 
\be 
\Delta T_{m,n} = g_{m,n}F(t_{n}^{(1)} - t_{m}^{(2)}).
\ee
where the function $F(x)$ is the analog of a phase response 
curve~\cite{Winfree80} for our model. To obtain results quantitatively 
comparable with our neuronal model, we fit it by non-negative 
quadratic function that describes phase response of our model
neurons: $F(x)=835+63x-9x^2$ for $0\le{x}\le{T_2^0}$ 
and 0 otherwise. $g_{m,n}$ obeys the inverse STDP update rules:
\bse \label{g-nm} \bea 
g_{m+1,n} & = & g_{m,n} - G(t_{m+1}^{(2)} - t_{n}^{(1)})
\\ g_{m,n} & = & g_{m,n-1} - G(t_{m}^{(2)} - t_{n}^{(1)}).
\eea \ese
%%%%%%%%%%%%%%%%%%%%%%%%%%%%%%%%%%%%%%%%%%%%%%%%%
\begin{figure}
  \begin{center}
\leavevmode
    \includegraphics[width=8.5 cm]{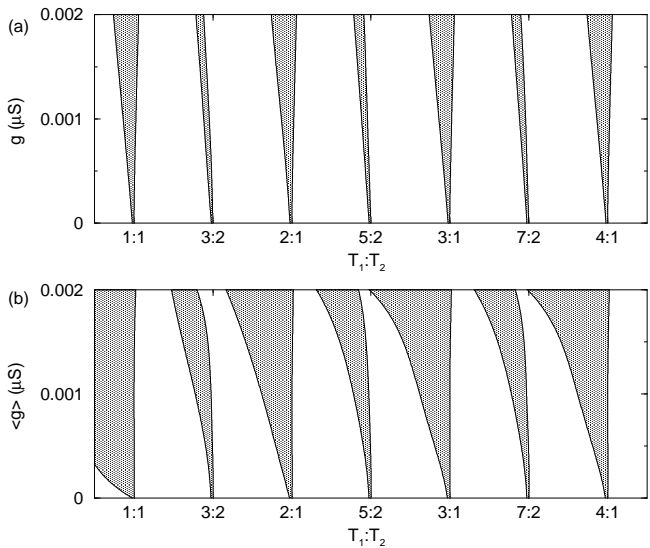}
  \end{center}
\caption{\label{AT_map}Arnol'd Tongues calculated for the discrete map
 model with (a) constant and (b) activity-dependent coupling. $T_2^0=13$ ms.}
\end{figure}
%%%%%%%%%%%%%%%%%%%%%%%%%%%%%%%%%%%%%%%%%%%%%%%%%%
In Fig.~\ref{AT_map} we show the Arnol'd Tongues 
calculated for the map (\ref{t-nm}-\ref{g-nm}) in the cases of (a) 
constant and (b) inverse STDP coupling. As with the model neurons,
we see that activity-dependent coupling greatly enlarges the zones
of synchronization.

This discrete map can be further analyzed to find its fixed points 
corresponding to $n:m$ synchronization and to examine their stability.
We present here only the case of 1:1 synchronization. Then $m=n$, and 
the system of equations (\ref{t-nm}-\ref{g-nm}) can be written in the
following simple form:
\bse \label{tau-n} \bea
\tau_{n+1} & = & \tau_{n} + T_1^0 - T_2^0 + g_{n}F(\tau_{n})
\\g_{n+1} & = & g_{n} - G(T_1^0-\tau_{n+1}) - G(-\tau_{n+1}),
\eea \ese
where $\tau_{n}=t_{n}^{(1)}-t_{n}^{(2)}$.
The fixed points of (\ref{tau-n}) are given by
$g_{n}^{f} = (T_2^0 - T_1^0)/F(\tau_{n}^{f})$ and $\tau_{n}^{f} = T_1^0/2$.
Stability calculations show that for such $F(\tau)$
and $G(\tau)$ these fixed points are stable. The second fixed point 
illustrates that activity-dependent coupling introduces a new limitation 
on the relationship between the phases of two oscillators. It is this
limitation that causes the suppression of phase slips under the influence
of noise. Detailed analysis shows that in the course of noise-affected 
synchronization the strength of activity-dependent coupling ajusts 
dynamically to keep this phase relationship close to satisfaction and,
hence, suppresses phase slips.

In conclusion, we have analyzed the effects of activity-dependent
coupling on synchronization properties of coupled neurons.
We showed that such coupling results in a substantial
extension of the temporal synchronization zones, leads to more rapid 
synchronization and makes it much more robust against 
noise. The enlargement of synchronization zones means that with 
STDP-like learning rules the number of synchronized neurons in a large
heterogeneous population must increase. In fact, this is an aspect of 
the popular idea due to Hebb~\cite{Hebb49}. It is supported by the 
results in~\cite{Miltner99, Fell01} which indicate that the coherence 
of fast EEG activity in the gamma band increases in a process of 
associative learning.

Based on our discrete map model results, we argue that the particular
details of the signal-generating devices (e.g. neurons) and their 
connections (e.g. synapses) are not essential and the obtained results
have general applicability. In fact, we observed similar phenomena of 
robust and enhanced synchronization in computer simulations of other 
types of periodic generators (such as Van-der-Pol and 
$\theta$-oscillators) with STDP-like activity-dependent coupling. 
\begin{acknowledgments}
This work was partially supported by U.S. Department of Energy 
Grants No. DE-FG03-90ER14138 and No. DE-FG03-96ER14592, NSF Grant 
No. PHY0097134, Army Research Office Grant No. DAAD19-01-1-0026, 
Office of Naval Research Grant No. N00014-00-1-0181, and NIH Grant 
No. R01 NS40110-01A2. R. Huerta thanks MCyT(Spain) BFI2000-0157.
\end{acknowledgments}


\begin{thebibliography}{18}
\bibitem[{par()}]{params}
\bibinfo{note}{The following values were used: 
  $g_{L}=0.027~\mu S,~E_{L}=-64~mV,~E_{Na}=50~mV,~g_{Na}=7.15~\mu S,
  ~g_{K}=1.43~\mu S,~E_{K}=-95~mV,
  ~\alpha = 10~ms^{-1},~\beta = 0.2~ms^{-1},~C=1.43\,10^{-4}~\mu F,
  ~\alpha_n =.032(-50-V)/(\exp((-50-V)/5)-1),~\beta_n=
  .5\exp((-55-V)/40),~\alpha_m=.32(-52-V)/(\exp((-52-V)/4)-1),
  ~\beta_m=.28(25+V)/(\exp((25+V)/5)-1),~\alpha_h=.128\exp((-48-V)/18)
  \text{ and } \beta_h=4/(\exp((-25-V)/5)+1)$}.
\end{thebibliography}
\end{document}